\documentclass[a4paper]{spie}  

\usepackage{amsmath,amsfonts,amssymb}
\usepackage{graphicx}
\usepackage[colorlinks=true, allcolors=blue]{hyperref}
\usepackage{siunitx}
\usepackage{subfigure}
\usepackage[colorinlistoftodos]{todonotes}
\usepackage{tikz}
\usepackage{verbatimbox}
\usepackage{pgfplots}
\usepackage{cleveref}
\usepackage{acro}                       
\DeclareAcronym{RMS}{
    short={RMS},
    long={root mean squared}
}

\DeclareAcronym{AO}{
    short={AO},
    long={adaptive optics}
}

\DeclareAcronym{XAO}{
    short={XAO},
    long={extreme adaptive optics}
}

\DeclareAcronym{SCAO}{
    short={SCAO},
    long={single conjugate adaptive optics}
}

\DeclareAcronym{DM}{
    short={DM},
    long={deformable mirror}
}

\DeclareAcronym{SLM}{
    short={SLM},
    long={spatial light modulator}
}

\DeclareAcronym{WFS}{
    short={WFS},
    long={wavefront sensor}
}

\DeclareAcronym{WFC}{
    short={WFC},
    long={wavefront controller}
}

\DeclareAcronym{PSF}{
    short={PSF},
    long={point spread function}
}

\DeclareAcronym{PSD}{
    short={PSD},
    long={power spectral density}
}

\DeclareAcronym{ESO}{
    short={ESO},
    long={European Southern Observatory}
}

\DeclareAcronym{FWHM}{
    short={FWHM},
    long={full width at half maximum}
}

\DeclareAcronym{KL}{
    short={KL},
    long={Karhunen-Lo\`{e}ve}
}

\DeclareAcronym{FOV}{
    short={FOV},
    long={field of view}
}

\DeclareAcronym{EELT}{
    short={the ELT},
    long={ESO's Extremely Large Telescope}
}

\DeclareAcronym{ELT}{
    short={ELT},
    long={extremely large telescope}
}
\DeclareAcronym{MAP}{
    short={MAP},
    long={\textit{maximum a posteriori}}
}

\DeclareAcronym{FFT}{
    short={FFT},
    long={fast Fourier transform}
}

\DeclareAcronym{DFT}{
    short={DFT},
    long={discrete Fourier transform}
}

\DeclareAcronym{GHOST}{
    short={GHOST},
    long={GPU-based High-order adaptive OpticS Testbench}
}
\DeclareAcronym{sLED}{
    short={sLED},
    long={super-luminous light-emitting diode}
}

\DeclareAcronym{PCS}{
    short={PCS},
    long={Planetary Camera and Spectrograph}
}

\DeclareAcronym{RTC}{
    short={RTC},
    long={Real-time computer}
}

\DeclareAcronym{GPU}{
    short={GPU},
    long={Graphics processing unit}
}
\DeclareAcronym{COTS}{
    short={COTS},
    long={Commercial off-the-shelf}
}

\DeclareAcronym{GUI}{
    short={GUI},
    long={graphical user interface}
}

\graphicspath{{./figures/}}

\title{The GPU-based High-order adaptive OpticS Testbench}

\author[a]{Byron Engler}
\author[a]{Markus Kasper}
\author[a]{Serban Leveratto}
\author[a]{Cedric Taissir Heritier}
\author[a]{Paul Bristow}
\author[a]{Christophe V\'erinaud}
\author[a]{Miska Le Louarn}
\author[b]{Jalo Nousiainen}
\author[b]{Tapio Helin}
\author[c]{Markus Bonse}
\author[c]{Sascha Quanz}
\author[c]{Adrian Glauser}
\author[d]{Julien Bernard}
\author[e]{Damien Gratadour}
\author[f]{Richard Clare}

\affil[a]{European Southern Observatory, Karl-Schwarzschild Stra\ss e 2, Garching bei Muenchen, Germany}
\affil[b]{LUT Lappeenranta}
\affil[c]{Eidgen\"ossische Technische Hochschule Z\"urich}
\affil[d]{Australian National University}
\affil[e]{Laboratoire d’\'etudes spatiales et d’instrumentation en astrophysique}
\affil[f]{Department of Electrical and Computer Engineering, University of Canterbury, New Zealand}

\authorinfo{Further author information: (Send correspondence to Byron Engler.)\\E-mail: byron.engler@eso.org}

\begin{document} 
\maketitle

\begin{abstract}
The GPU-based High-order adaptive OpticS Testbench (GHOST) at the European Southern Observatory (ESO) is a new 2-stage extreme adaptive optics (XAO) testbench at ESO. The GHOST is designed to investigate and evaluate new control methods (machine learning, predictive control) for XAO which will be required for instruments such as the Planetary Camera and Spectrograph of ESOs Extremely Large Telescope. The first stage corrections are performed in simulation, with the residual wavefront error at each iteration saved. The residual wavefront errors from the first stage are then injected into the GHOST using a spatial light modulator. The second stage correction is made with a Boston Michromachines Corporation 492 actuator deformable mirror and a pyramid wavefront sensor. The flexibility of the bench also opens it up to other applications, one such application is investigating the flip-flop modulation method for the pyramid wavefront sensor.

\end{abstract}

\keywords{wavefront sensing, adaptive optics, pyramid wavefront sensor, EELT, petaling, segment piston, predictive control, machine learning, extreme adaptive optics, PCS, GHOST}

\section{INTRODUCTION}
    \label{sec:intro}  
    The \ac{PCS} for the \ac{EELT} will focus on the direct imaging and characterising of nearby exoplanets \cite{pcs2021}. The \ac{PCS} will combine \ac{XAO}, coronagraphy (high contrast imaging) and spectroscopy to achieve extremely high contrast between the on-axis starlight and the off-axis light reflected by the planet. The \ac{XAO} of the \ac{PCS} will consist of two stages, a slow (\SI{1}{\kilo\hertz}) loop controlling the first-stage \ac{DM} with approximately 4000 modes, and a high-speed (\SI{4}{\kilo\hertz}) loop controlling the second-stage \ac{DM} with on the order of 10000 modes. Predictive control methods, which use past measurements to make predictions of the wavefront at the current time, have an advantage over the traditional integrator controller. A predictive controller will reduce the temporal error and could also reduce the impact of photon noise \cite{males_predictive}. Another potential control method would be to control the \ac{PSF} contrast directly, instead of directly flattening the wavefront. A promising field of research is using machine learning techniques for both predictive control and \ac{PSF} contrast control. 
    
    The \ac{ESO}, along with collaborations from external institutes (ETH Zurich, LUT Lappeenranta) has developed the \ac{GHOST} as a tool to facilitate the development of predictive control techniques with machine learning in the context of \ac{XAO} for the \ac{PCS}. The \ac{GHOST} is a two-stage \ac{XAO} system, dimensioned to be similar to the VLT/SPHERE. The first-stage control is performed in simulation, using either a pyramid or Shack-Hartmann wavefront sensor with 40x40 sub-apertures, typically controlling 800-1000 modes at \SI{1}{\kilo\hertz}. The residual wavefront error is saved and injected onto the bench using a \ac{SLM}. The second-stage \ac{AO} control is performed using a pyramid wavefront sensor and a Boston Micromachines Corporation \ac{DM}, running at an effective speed of \SIrange{2}{4}{\kilo\hertz}, and controlling 300-400 modes. The \ac{RTC} is built using \ac{COTS} server components and two Nvidia enthusiast gaming \ac{GPU}s, which will be used for the real-time computation.



\section{Design and implementation}
\label{sec:lab_setup}
\begin{figure}
    \centering
        \includegraphics[width=1\textwidth, clip, trim=0.5cm 2cm 2cm 0.5cm]{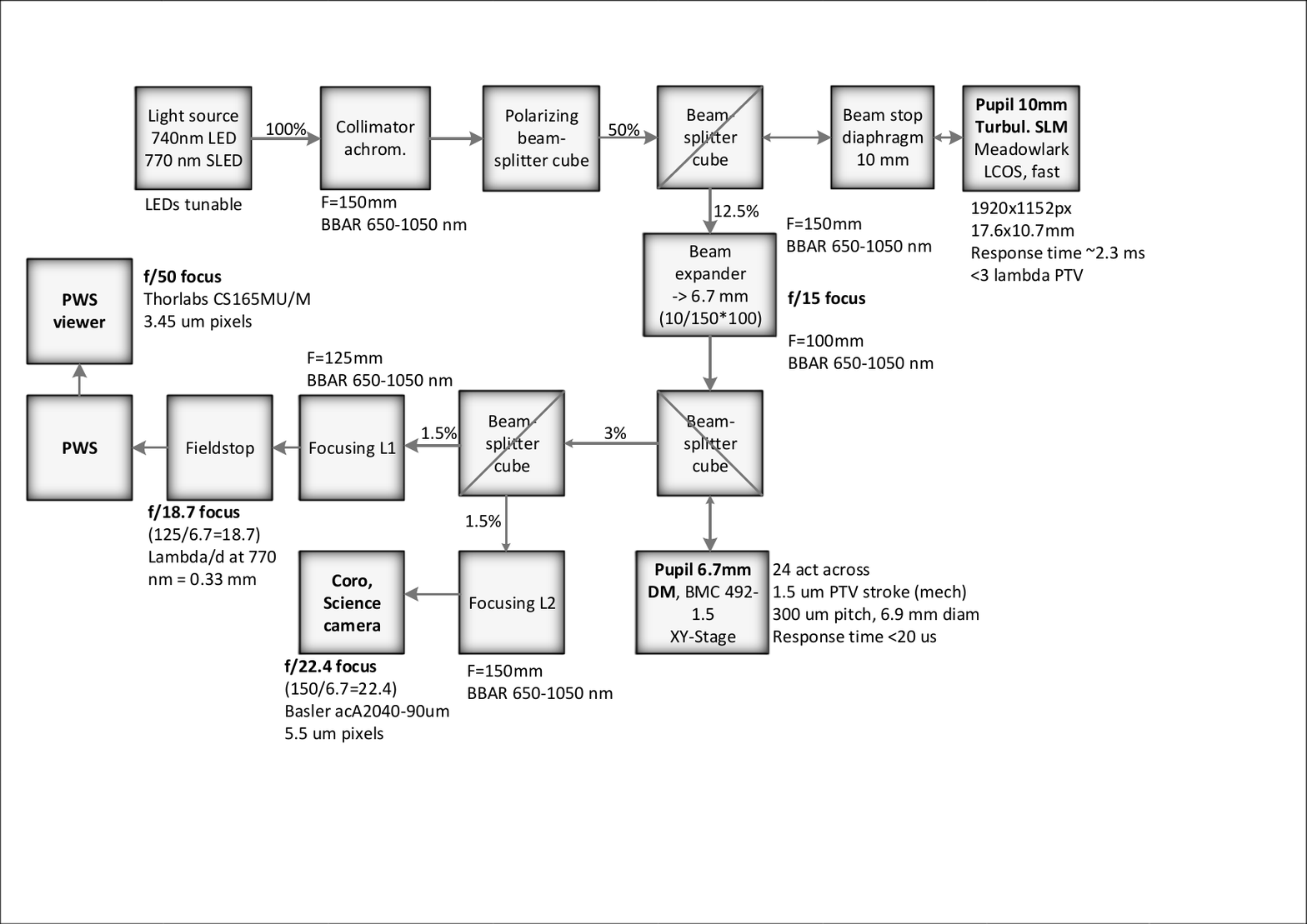}
        \caption{A schematic of the \ac{GHOST} optics. The light source throughput, beam size and focal ratio are noted at key locations.}\label{fig:optics}     
\end{figure}

\begin{figure}
    \centering
        \includegraphics[width=1\textwidth]{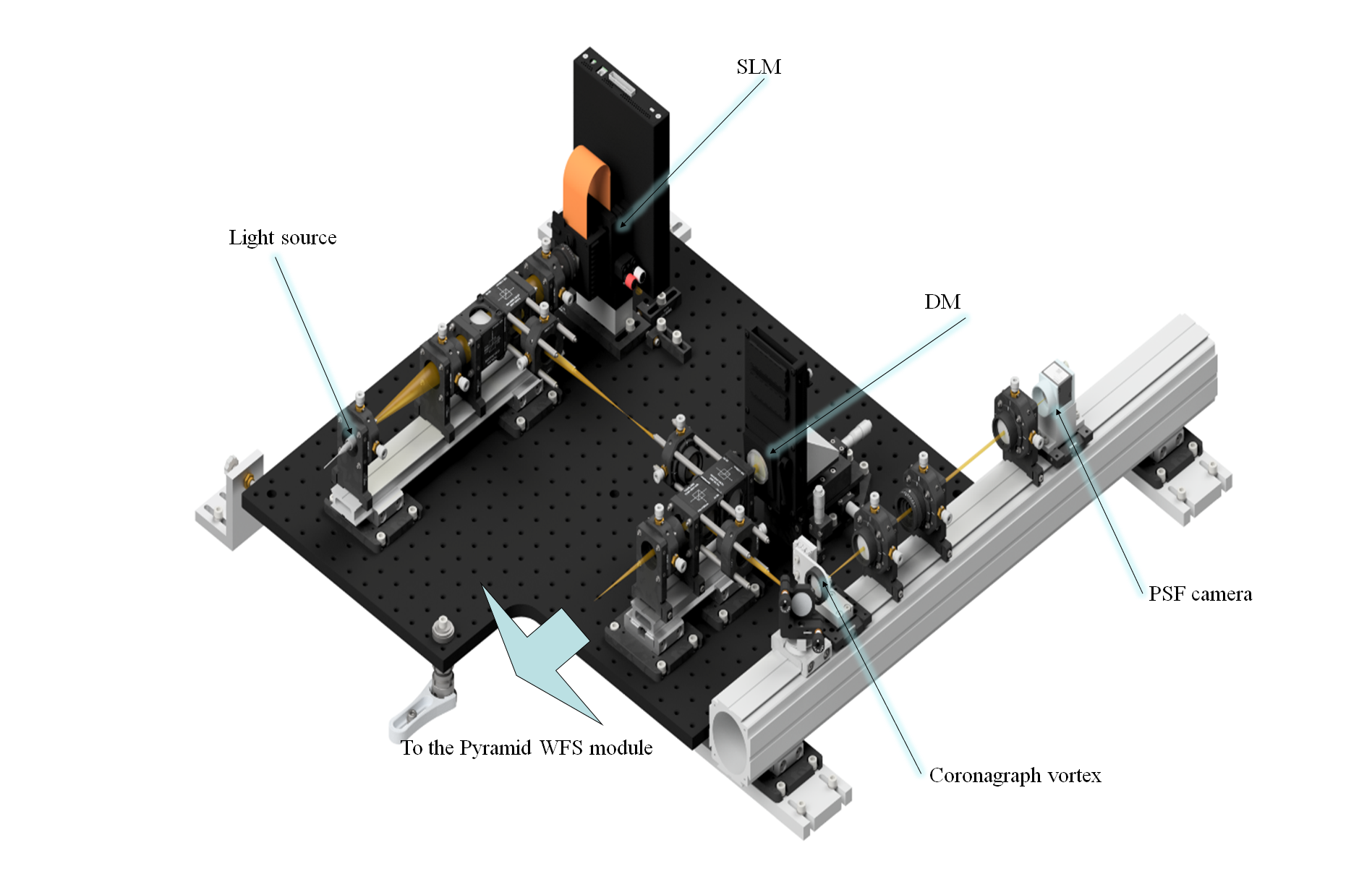}
        \caption{A 3D render of the main optical components of the GHOST bench, showing the light path through the system. The pyramid module is recycled from the High Order Testbench.}\label{fig:render}     
\end{figure}

The \ac{GHOST} optical design is relatively simple, using on-axis optics and beams splitters and is broken into three separate modules. A schematic of the optical configuration is shown in \cref{fig:optics} and a 3D render of the first two modules is shown in \cref{fig:render}. The simplicity and modularity of the design allows for straightforward optical alignment, each module can be independently aligned off of the main optical bench. The trade-off with this design is that only 1.5\% of the input light makes it to the \ac{WFS}, this is offset by having a powerful light source. The modules are as follows:
\begin{itemize}
    \item Light source, Spatial Light Modulator (SLM), and \ac{DM}.
    \item Coronagraph and PSF imager (‘science' camera).
    \item Pyramid \ac{WFS}, modulation mirror, and pyramid viewing camera. The pyramid \ac{WFS} has been developed by Arcetri and provided to ESO in the frame of an OPTICON collaboration in 2006.
 The \ac{GHOST} is also fully remotely controllable, which proved to be extremely useful during COVID-19 restrictions but also makes the GHOST accessible to anyone with an internet connection. 
\end{itemize}

\subsection{Light source}
    The light source for the \ac{GHOST} bench is a single-mode fibre-coupled \ac{sLED}, sourced from ThorLabs (model number: SLD770S). The \ac{sLED} has a nominal wavelength of \SI{770}{\nano \metre} with a \SI{18}{\nano \metre} bandwidth, and an output power of \SI{5.5}{\milli \watt}. The \ac{sLED} is driven with a ThorLabs CLD1015 compact laser diode driver. The CLD1015 controls the current through the LED and the temperature of the diode. The CLD1015 also allows remote control of the light source via a USB connection to a computer.
    
    \begin{table}
        \centering
        \bigskip
        \caption{\label{table:light_source} \ac{sLED} specifications}
        \begin{tabular}{cc}
            \hline\noalign{\smallskip} Parameter  & Value  \\
            \noalign{\smallskip}\hline\noalign{\smallskip} 
            Model & Thorlabs SLD770S\\
            Centre wavelength & \SI{770}{\nano\metre}  \\
            Bandwidth & \SI{18}{\nano\metre} \\
            Fiber-coupled output power & \SI{5.5}{\milli\watt}\\
            Driver & Thorlabs CLD1015
        \end{tabular}
    \end{table}

\subsection{Spatial light modulator}
    The key specification for the \ac{SLM} selected for the \ac{GHOST} is the maximum frame rate. The selected \ac{SLM}, Meadowlark  HSP1920-600-1300-HSP8, has a maximum refresh rate of \SI{422.4}{\hertz}. The key specifications for the \ac{SLM} are listed in \cref{table:slm}. The \ac{SLM} is provided with a lookup table (LUT), which maps the commanded value (0-255) to phase, for \SI{1550}{\nano\metre}. The provided LUT mapped $0-255$ to $0-2\pi$ radians of phase shift. Initially, we simply scaled the LUT to the wavelength of our light source. Using the calibration tools provided by Meadowlark, a custom LUT was measured with the \ac{SLM} in the \ac{GHOST}, with a range of $0-2\pi$ and $0-4\pi$. The $0-2\pi$ LUT gives a smaller phase step size, while the $0-4\pi$ gives a larger maximum phase shift. 
    \begin{table}
        \centering
        \bigskip
        \caption{\label{table:slm} \ac{SLM} specifications}
        \begin{tabular}{cc}
            \hline\noalign{\smallskip} Parameter  & Value  \\
            \noalign{\smallskip}\hline\noalign{\smallskip} 
            Model               & Meadowlark  HSP1920-600-1300-HSP8 \\
            Maximum frame rate  & \SI{422.4}{\hertz}, optical settling time 2.3ms \\
            Resolution          & 1920 x 1152 pixels                \\
            Pixel size         & \SI{9.2}{\micro\metre} square     \\
            Fill factor         & 95.7\%                            \\
            Zero-order diffraction efficiency & 88\%                \\
            Control             & 8-bit, 256 grayscale values       \\
            Stroke              & $4\pi$ at \SI{770}{\nano\metre}
            
        \end{tabular}
    \end{table}

\subsection{Deformable mirror}
The deformable mirror is a Boston Micromachines Corporation DM 492-C, which is on loan to \ac{ESO} from ETH Zurich for the duration of the project. The key parameters of the \ac{DM} are listed in \cref{table:dm}. The stroke of the \ac{DM} is not suitable for correction open loop atmospheric turbulence, however, for the application in the \ac{GHOST} it is ideal as it effectively operates as a second stage \ac{DM}. The \ac{DM} driver electronics connects to the RTC via a fibre optic cable connection to a PCIe card provided by Boston Micromachines Corporation.
    \begin{table}
        \centering
        \bigskip
        \caption{\label{table:dm} \ac{DM} specifications}
        \begin{tabular}{cc}
            \hline\noalign{\smallskip} Parameter  & Value  \\
            \noalign{\smallskip}\hline\noalign{\smallskip} 
            Model               & Boston Micromachines Corporation DM 492-C \\
            Actuator grid       & 24 x 24   \\
            Actuator pitch      & \SI{300}{\micro\metre} \\
            Active aperture size &  \SI{6.9}{\milli\metre} \\
            Available actuators & 492 \\
            Stroke              & \SI{1500}{\nano\metre} mechanical deflection
                    
        \end{tabular}
    \end{table}

\subsection{Coronagraph and imaging camera}
\begin{figure}
    \centering
        \includegraphics[width=0.8\textwidth]{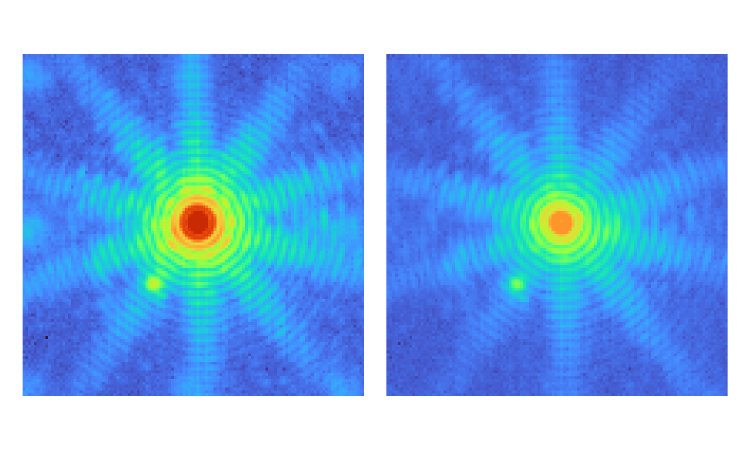}
        \caption{The PSF of the \ac{GHOST} with the vortex coronagraph moved out of the optical path (left) and in the optical path (right). The speckle in the lower left of the images is due to a reflection (ghost) in the optics.\label{fig:coro} }
\end{figure}

The imaging camera for the \ac{GHOST} is a Basler ACA2040-90um, the specifications are listed in \cref{table:psf_cam}. The camera arm consists of a vortex coronagraph and the associated relay optics. The camera can be placed in either the focal plane (to image the PSF of the system) or the pupil plane (to adjust the Lyot stop of the coronagraph). The vortex phase plate of the coronagraph is on a motorised XY stage and can be moved in or out of the optical path as required. \Cref{fig:coro} shows the PSF of the \ac{GHOST}, with no turbulence (near-diffraction limited), with and without the coronagraph in place. There is a notable reduction in flux when the coronagraph vortex is in the optical path, however, the reduction is only 2-3 times which is significantly less than desired. Upon investigation, the vortex phase plate was found to have a defect where it has two sweet spots instead of one. We are working with the manufacturer to resolve the issue.
    \begin{table}
        \centering
        \bigskip
        \caption{\label{table:psf_cam} PSF/coronagraph camera}
        \begin{tabular}{cc}
            \hline\noalign{\smallskip} Parameter  & Value  \\
            \noalign{\smallskip}\hline\noalign{\smallskip} 
            Model               & Basler ACA2040-90um \\
            Resolution       & 2048 x 2048 pixels   \\
            Pixel size      & \SI{5.5}{\micro\metre} square \\
            PSF size ($\frac{\lambda}{D}$) & 4 pixels\\
            Vortex mask & Thorlabs WPV10-780
        \end{tabular}
    \end{table}

\subsection{Pyramid wavefront sensor}
\begin{figure}
    \centering
        \includegraphics[width=0.7\textwidth]{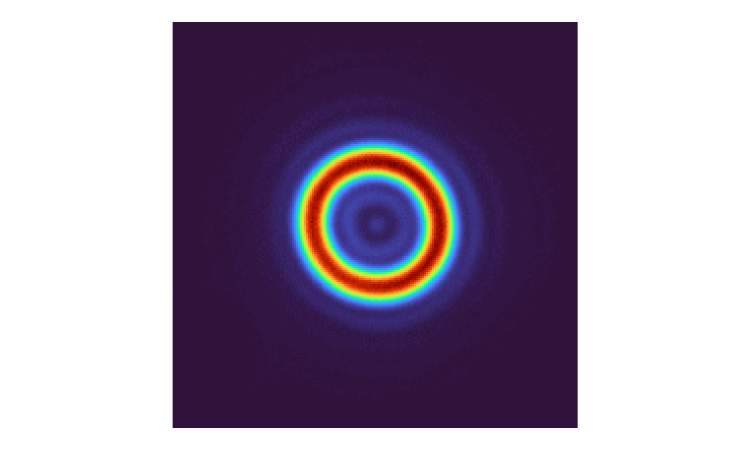}
        \caption{The circularised modulation path, with a radius of $\frac{3\lambda}{D}$ and a modulation frequency of \SI{400}{\hertz}, as seen by the pyramid viewing camera.\label{fig:pyr_viewer} }
\end{figure}

\begin{figure}
    \centering
        \includegraphics[width=0.8\textwidth]{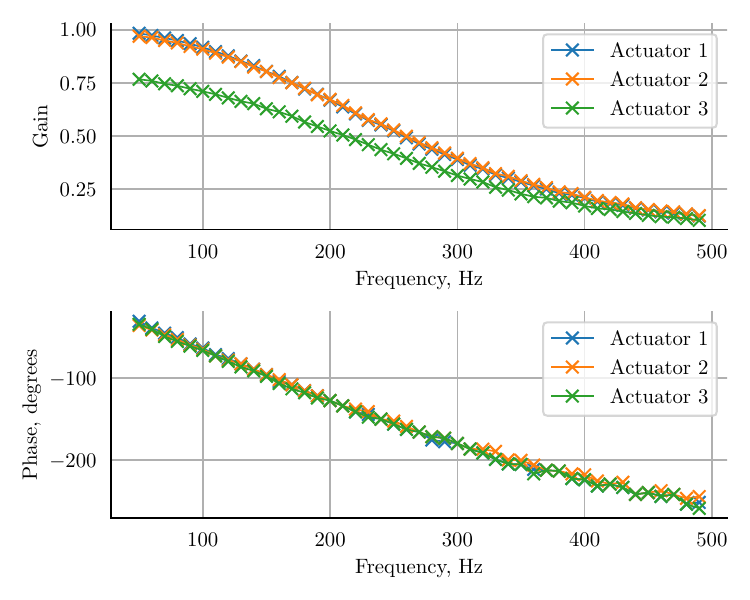}
        \caption{The closed-loop gain and phase response of each actuator whilst modulation at a range of frequencies. The measurements are made using the strain gauges built into each actuator of the modulation mirror.\label{fig:mod_mirror_bode} }
\end{figure}

The pyramid \ac{WFS} module has been developed by Arcetri and provided to ESO in the frame of an OPTICON collaboration in 2006, for use on the High Order Testbench (HOT). This module has been recycled, with an upgraded camera. The specifications of the \ac{WFS} camera are listed in \cref{table:wfs_cam}.
\begin{table}
    \centering
    \bigskip
    \caption{\label{table:wfs_cam} Pyramid wavefront sensor camera}
    \begin{tabular}{cc}
        \hline\noalign{\smallskip} Parameter  & Value  \\
        \noalign{\smallskip}\hline\noalign{\smallskip} 
        Model               & Baumer VLXT-06M.I \\
        Resolution       & 800 x 620 pixels   \\
        Pixel size      & \SI{9.9}{\micro\metre} square \\
        Maximum frame rate (full frame) & 1578 fps \\
        PWFS pupil diameter & 36 pixels
    \end{tabular}
\end{table}
The modulation tip/tilt mirror of the pyramid \ac{WFS} is produced by Physik Instrumente, and the key specifications are listed in \cref{table:mod_mirror}. The mirror is actuated by 3 actuators, each with a separation angle of \SI{120}{\degree}. In order to accurately produce the commanded modulation pattern, the gain and phase delay of each actuator is first measured using the position sensors built into the actuators. \Cref{fig:mod_mirror_bode} shows the gain and phase response of the modulation mirror for a range of modulation frequencies. Knowing the gain and phase delay for each actuator, a suitable command can be found such that the modulation path is circular. However, due to the optical configuration, the gain on one axis is less than on the other, without an optical calibration the modulation path would not be circular. For the purpose of optically calibrating the modulation path, a pyramid viewing camera was added to the pyramid WFS module. A pellicle beamsplitter is inserted just in before the pyramid which diverts light to a pyramid viewing camera, effectively measuring the focal plane in which the pyramid is placed. This allows the modulation mirror to be calibrated such that the modulation radius is known in $\frac{\lambda}{D}$, as well as the optical gain for each of the actuators. \Cref{fig:pyr_viewer} shows the modulation path as seen by the pyramid viewing camera, with a corrected circular modulation.

\begin{table}
    \centering
    \bigskip
    \caption{\label{table:mod_mirror} Pyramid modulation mirror}
    \begin{tabular}{cc}
        \hline\noalign{\smallskip} Parameter  & Value  \\
        \noalign{\smallskip}\hline\noalign{\smallskip} 
        Model               & Physik Instrumente SL325 \\
        Controller & E-518\\
        Mechanical Resonant Frequency & \SI{1}{\kilo\hertz} loaded (\SI{2}{\kilo\hertz} unloaded)
    \end{tabular}
\end{table}

\subsection{Real-time computer}
\begin{figure}
    \centering
        \includegraphics[width=1.25\textwidth, angle=90]{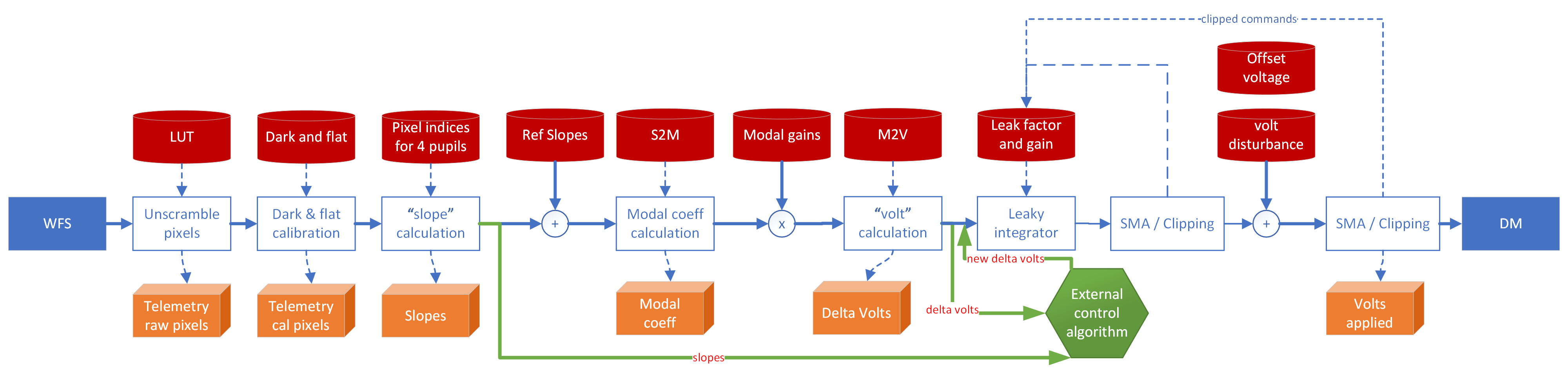}
        \caption{The real-time pipeline implemented using the COSMIC RTC.\label{fig:RTC} }
\end{figure}

The real-time computer consists of \ac{COTS} server components, with two Nvidia gaming \ac{GPU}s. The major components of the computer are listed in \cref{table:RTC}.
The \ac{GHOST} will use the COSMIC RTC platform, making use of the \ac{GPU}s in the computer. COSMIC is a hard real-time adaptive optics controller, implemented on Nvidia \ac{GPU}s \cite{cosmic}. Partnering with the Australian National University and LESIA, the COSMIC platform has been installed on the \ac{GHOST}. \Cref{fig:RTC} shows the COSMIC pipeline implemented for the \ac{GHOST}.
\begin{table}
    \centering
    \bigskip
    \caption{\label{table:RTC} Real-time computer}
    \begin{tabular}{cc}
        \hline\noalign{\smallskip} Parameter  & Value  \\
        \noalign{\smallskip}\hline\noalign{\smallskip} 
        CPU                 & 2x Intel(R) Xeon(R) Gold 6258R  (112 threads total) \\
        RAM                 & 192 GB DDR4 (12 x 16GB) \\
        GPU                 & 2x Nvidia RTX Titan with NVlink bridge  
    \end{tabular}
\end{table}

\section{Adaptive optics calibration}
\begin{figure}
    \centering
        \includegraphics[width=1\textwidth]{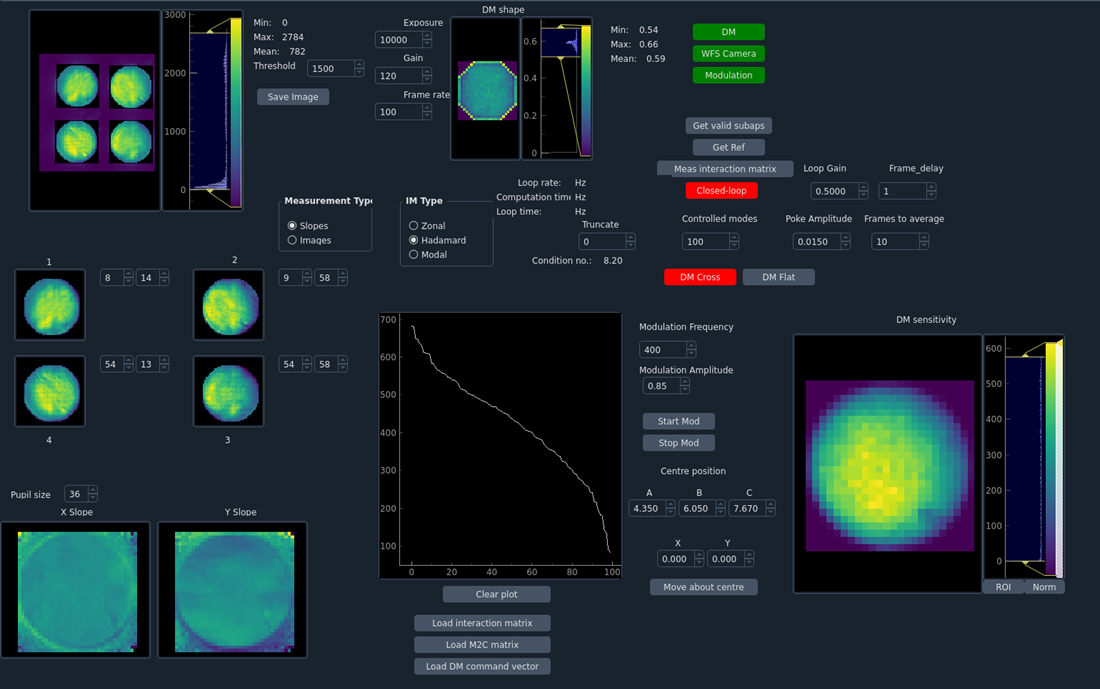}
        \caption{The graphical user interface developed in Python using PyQt5 and PyQtGraph, for the alignment and adaptive optics calibration, and closed-loop control. The plot at the top left is the latest image from the wavefront sensor. The four pupils are masked, and the cut-outs are shown immediately below. Below the pupil cutouts are the x and y slopes. The plot in the centre shows the eigenvalues of the control matrix. The figure at the top right shows the current DM command. The plot at the bottom right shows the sensitivity of the DM actuators in the WFS space.\label{fig:gui} }
\end{figure}

For the prupose of commissioning and adaptive optics calibration, a \ac{GUI} has been developed using Python. The NumPy library, compiled with the Intel MKL BLAS and LAPACK, is used for all of the computation, PyQt5 is used for the \ac{GUI} and PyQtGraph is used for the real-time figures. The \ac{GUI} is shown in \cref{fig:gui}, after completion of the \ac{AO} calibration. The \ac{GUI} is also capable of closed-loop \ac{AO} control.

The calibration of the \ac{AO} system is achieved by actuating the \ac{DM} with Hadamard patterns, and the resulting \ac{WFS} signals are decoded and inserted into a zonal interaction matrix \cite{hadamard}. Mathematically, the calibration procedure is for a system with $n$ actuators and $m$ slope measurements, expressed as
\begin{equation}
    C = DV + N,
\end{equation}
where $C$ is an $m$x$n$ matrix containing \ac{WFS} slope measurements, V is an $n$x$n$ matrix containing the actuation patterns, and $D$ is an $m$x$n$ and is the interaction matrix relating slopes to actuation patterns. An estimation of the interaction matrix, $D$, is then
\begin{equation}
    \hat{D} = CV^{-1}.
\end{equation}
A Hadamard matrix is an orthogonal square matrix containing only 1's and -1's. The inverse of a Hadamard matrix of size $n$, $H_n$ is then
\begin{equation}
    H_n^{-1} = H_n^T.
\end{equation}
If we actuate the \ac{DM} with Hadamard patterns,
\begin{equation}
    V = v_m H_n,
\end{equation}
where $v_m$ is a scalar, then the interaction matrix can be estimated as
\begin{equation}
    \hat{D} = C V^T.
\end{equation}
In the case of the \ac{GHOST}, the nearest Hadamard matrix has a size of $n=512$. In this case, the \ac{DM} commands are truncated to the number of actuators (492)
For modal control, the modal basis is then projected onto the zonal interaction matrix, giving a modal interaction matrix. The pseudo-inverse of the modal interaction matrix is then used as the control matrix.

\section{Current status and applications of \ac{GHOST}}
\begin{figure}
    \centering
        \includegraphics[width=1\textwidth,clip, trim=0.1cm 2cm 0.1cm 2cm]{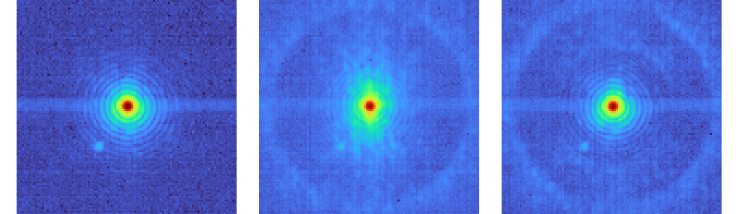}
        \caption{The long exposure point spread functions measured with the focal plane imager. From left to right: The diffraction PSF of the system, the PSF with AO residuals from the simulated first stage where the control radius of the first stage is visible (900 modes controlled), and the PSF with the 2nd stage loop closed where the control radius of the second stage is clearly visible, with 350 controlled modes. The speckle in the upper left quadrant is a reflection (ghost) in the system.\label{fig:psfs} }
\end{figure}

\begin{figure}
    \centering
        \includegraphics[width=1\textwidth]{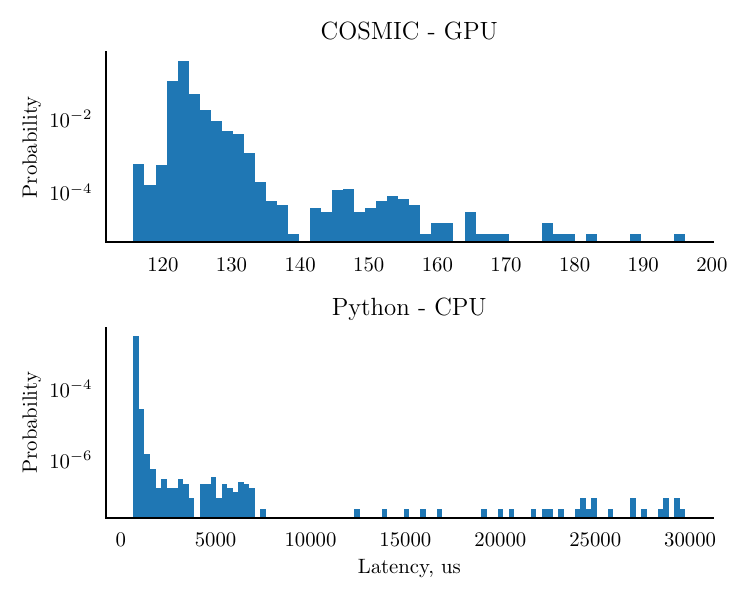}
        \caption{The latency and jitter from WFS image received to DM commands sent for the COSMIC RTC and the Python \ac{AO} CPU implementation.\label{fig:RTC_timing} }
\end{figure}

The \ac{GHOST} is fully operational, \cref{fig:psfs} shows the long exposure PSFs measured with the PSF camera for the diffraction-limited case (no turbulence), the open-loop (1\textsuperscript{st} stage closed-loop residuals) case, and the closed-loop (2\textsuperscript{nd} stage closed) case.
The computation time for the closed-loop calculations (time from receiving the image to sending the DM commands) for both the COSMIC GPU RTC and Python-based commissioning GUI are shown in \cref{fig:RTC_timing} whilst controlling 300 modes. The GPU-based COSMIC RTC achieves an average latency of \SI{123}{\micro\second} whilst the CPU-based Python GUI archives an average latency of \SI{964}{\micro\second}. No special effort has been made to optimise the Python GUI and as such there are large spikes in latency.
The \ac{GHOST} is already in use, with three papers presented at the SPIE Astronomical Telescopes + Instrumentation 2022 conference making use of the \ac{GHOST}:
\begin{itemize}
    \item Advances in model-based reinforcement learning for AO control – Jalo Nousiainen - Paper 12185-302
    \item Detection of discontinuous phase steps with a pyramid wavefront sensor – Deborah Malone - Paper 12185-200
    \item Experimental verification of a Neural Network and PCA approach for NCPA mitigation - Alessandro Terreri - Paper 12185-314
\end{itemize}

\begin{figure}
    \centering
        \includegraphics[width=1\textwidth]{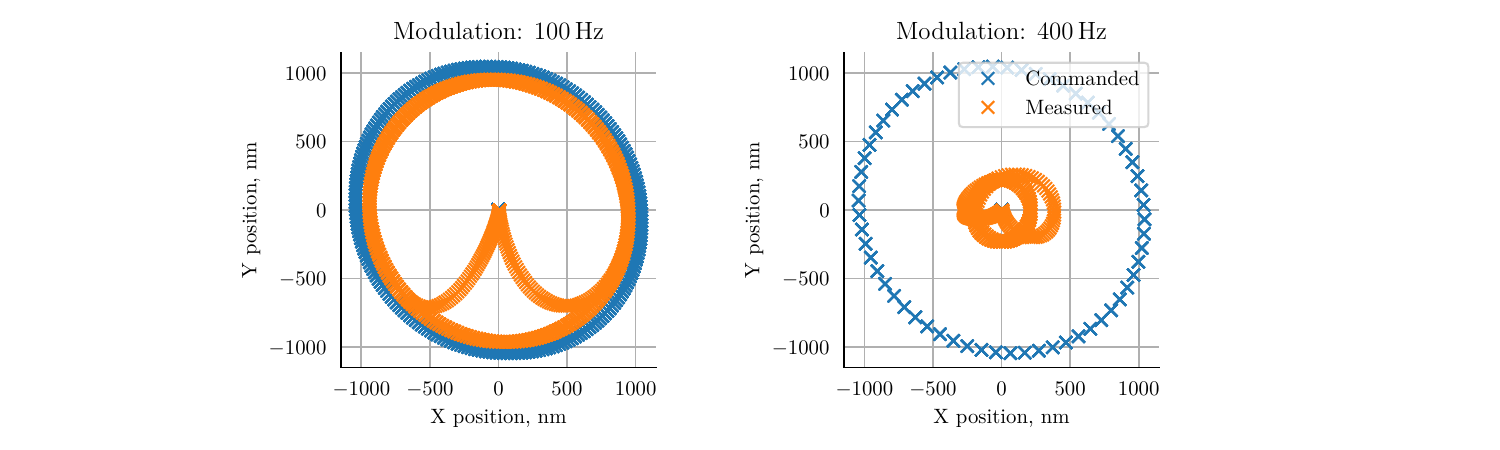}
        \caption{The commanded and measured modulation path at 100Hz (left) and 400Hz (right). The commanded paths have been corrected to produce a circle at the tip of the pyramid. The actuators of the modulation mirror have a frequency-dependent gain and phase shift, which need to be considered for each modulation frequency. The modulator is clearly able to start and stop within the one frame required for flip-flop modulation at both 100Hz and 400Hz.\label{fig:flip_flop_modulation}}
\end{figure}

Whilst the GHOST is intended to explore new control methods for XAO, it is fully equipped for other experiments which require a pyramid wavefront sensor, a high order DM, and a method to inject arbitrary phase screens. One such application is exploring flip-flop modulation \cite{engler22}, where the pyramid wavefront sensor is operated in both a modulated and unmodulated state. A key question posed by the flip-flop modulation method is whether the modulator could be stopped and started rapidly. \Cref{fig:flip_flop_modulation} shows the commanded and measured modulation path at 100Hz and 400Hz. In both cases, the modulator can start and stop within a single modulation cycle. 

\begin{figure}
    \centering
        \includegraphics[width=0.8\textwidth]{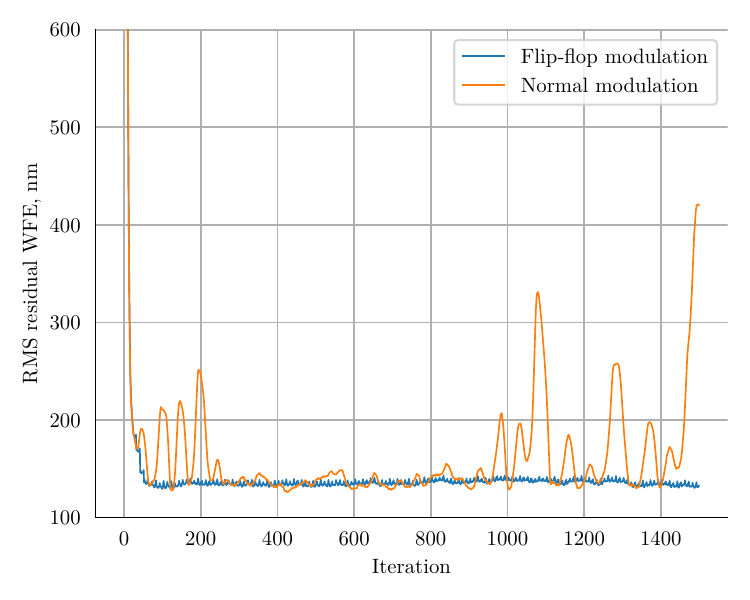}
        \caption{A simulation comparing the flip-flop modulation method (orange) with ‘normal’ modulation (blue). The root mean square of the residual wavefront error is plotted at each iteration. The atmosphere has an $r_0$ of \SI{15}{\centi\metre} and the wavefront sensor is in R-band.\label{fig:flip_flop_comparison}}
\end{figure}

\begin{figure}
    \centering
        \includegraphics[width=0.8\textwidth]{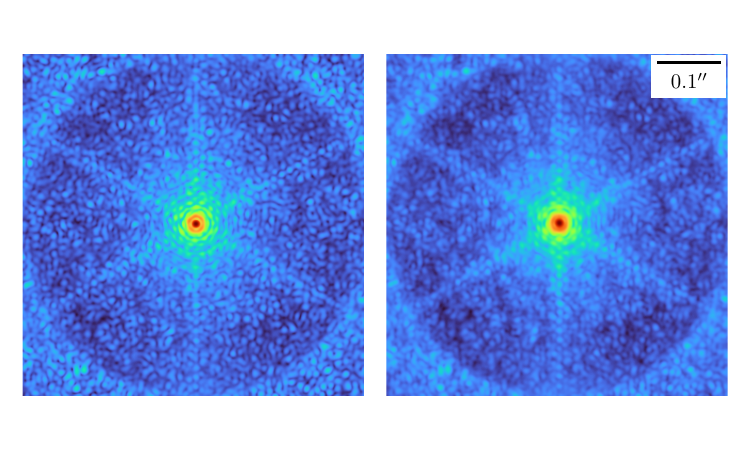}
        \caption{The simulated long exposure point spread function in R-band, with the WFS operating in R-band, with flip-flop modulation (left) and normal modulation (right). \label{fig:flip_flop_psf}}
\end{figure}

Using the parameters in \cref{tab:sim}, the flip-flop modulation is compared to a normal circular modulation pattern. \Cref{fig:flip_flop_comparison} compares the root mean square of the residual wavefront error and the resulting long exposure PSFs are shown in \cref{fig:flip_flop_psf}. For the normal modulation case, a continuous modal basis \cite{Bertrou-Cantou} is used. To create the continuous basis used, the \ac{KL} modes are defined on a circular pupil geometry, without spiders. The flip-flop modulation uses petal eigenmodes for the unmodulated pyramid, and \ac{KL} modes defined with the pupil geometry (with spiders) and forced to be orthogonal to the petal eigenmodes are used for the modulated pyramid.

\begin{table}[tb]
  \centering
  \bigskip
  \caption{\label{tab:sim} Simulation parameters used for flip-flop modulation.}
  
    \begin{tabular}{cc}
        \hline\noalign{\smallskip} Parameter  & Value  \\
        \noalign{\smallskip}\hline\noalign{\smallskip} 
        Telescope Diameter (D) & \SI{40}{\meter} \\
        Secondary Obstruction & 28\%\\
        Fried Parameter ($r_0$) &  \SI{15}{\centi\meter}\\
        Outer Scale ($L_0$)& \SI{30}{\meter} \\
        Atmosphere & single layer  \\
        Frame Rate ($F_s$) & \SI{1}{\kilo \hertz} \\
        Delay & 2 Frames \\
        Controller Type & Integrator\\
        PSF ($\lambda_{p}$ ) Wavelength &
        \SI{640}{\nano \meter}
        (R-band) \\
        WFS  ($\lambda_{W}$) Wavelength &
         \SI{640}{\nano \meter}
        (R-band) \\
        WFS Order & 116 $\times$ 96 Subapertures \\
        Modulation Width &  $4\lambda_W / D$,  $0\lambda_W / D$ \\
        Time Steps & 1500 \\
        Number of Spider Arms & 6 \\
        Spider Arm Width & \SI{51}{\centi\meter} \\ 
        Flux & 10000 photons/subaperture/frame\\
        Number of Actuators & 5190\\
        Number of Modes & 3500 \\
        \noalign{\smallskip}\hline
    \end{tabular}
\end{table}

\section{CONCLUSION}
\label{sec:conclusion}
This paper presents the design and implementation of the \ac{GHOST} which enables the development of predictive control in the context of \ac{XAO} for PCS. The \ac{GHOST} is a 2-stage \ac{XAO} system, where the first stage is simulated, with first stage residual turbulence injected to the bench with a high-speed \ac{SLM}. The \ac{GHOST} is fully operational and being actively used, the first stage (simulated and injected onto the bench with the \ac{SLM}) controls 800-1000 modes whilst the second stage controls 300-400 modes, at a loop rate of up to \SI{400}{\hertz} (emulating a loop rate of up to \SI{4}{\kilo\hertz}). Whilst \ac{GHOST} has been designed to develop predictive control algorithms, it also provides a general purpose \ac{AO} testbed, which is remotely accessible, controllable via Python interfaces and easy to implement new control ideas on.

\section{Acknowledgements}
The authors would like to thank ESO's technology development programme, which has helped fund the development of the \ac{GHOST}.
\bibliography{report} 
\bibliographystyle{spiebib} 

\end{document}